\documentstyle[12pt]{article}

\newcommand{\beq}{\begin{equation}}
\newcommand{\eeqn}{\end{eqnarray}}
\newcommand{\beqn}{\begin{eqnarray}}
\newcommand{\eeq}{\end{equation}}
\newcommand{\np}{Nucl. Phys. \underline}    
\newcommand{\pl}{Phys. Lett. \underline}    
\newcommand{\pr}{Phys. Rev. \underline}     

\title{Computing the Slope of the Isgur-Wise Function}
\author{U.Aglietti$^1$, G.Martinelli$^1$ and
C.T.Sachrajda$^2$\\  \mbox{}\\  $^1$Dipartimento di
Fisica, Universit\` a di Roma \lq La Sapienza\rq, \\ I-00185 Roma,
Italy \\ and\\ INFN, Sezione di Roma, Italy, \\ \\ $^2$Physics
Department, University of Southampton, \\ Southampton SO9 5NH, UK, \\ }
\date{}
\begin{document}
\maketitle
\begin{abstract}
We propose a method for evaluating the slope (and higher derivatives)
of the Isgur-Wise function at the zero recoil point using lattice simulations.
These derivatives are required for the extrapolation of the experimental
data for $B\rightarrow D^*l\bar\nu$ decays to the zero recoil point,
from which the $V_{cb}$ element of the CKM-matrix can be determined.
\vspace{-6.6truein}
\flushright{Southampton Preprint SHEP 93/94-07}\\
\flushright{Rome Preprint 93/983 }
\end{abstract}
\vspace{6truein}
The ARGUS \cite{argus} and CLEO \cite{cleo}
collaborations study the  semi-leptonic decays $B\rightarrow
D^*l\bar\nu$, and present results for  $|V_{cb}|\xi(\omega)$, where
$\xi(\omega)$ is the Isgur-Wise function \cite{iw}, $\omega$ is the dot
product of the four-velocities of the $B$ and $D^*$ mesons ($\omega =
v_B\cdot v_{D^*}$), and $V_{cb}$ is the $cb$ element of the
Cabibbo-Kobayashi-Maskawa matrix. The  presentation of the results in
this form is suggested by the ``Heavy Quark Effective Theory" (HQET)
(for a comprehensive review and references to the  original literature
see ref.\cite{neubert}). The Isgur-Wise function is normalised to 1 at
the zero recoil point, i.e. $\xi(1)=1$, so the extraction of the
$V_{cb}$ matrix element requires an extrapolation of the experimental
results from $\omega > 1$ to $\omega=1$
\footnote{In principle $V_{cb}$ can also be determined by
fitting the data to a theoretically predicted function of $\omega$, however
we wish to make strong use of the rigorous result $\xi(1)=1$.}.
This extrapolation requires a
model function or a theoretical prediction as a guide. Recently, the
first lattice computations of $\xi(\omega)$ have been performed
(for discrete values of $\omega>1$ \cite{bss,ukqcd},
and  have been used to extrapolate the  experimental
data to $\omega=1$, and hence to determine $V_{cb}$. In this letter we
point out that it is also possible to compute directly the slope of the
Isgur-Wise function at the zero-recoil point ($\xi^\prime(1)$), as well
as some higher derivatives.

The calculations suggested in this letter complement those already
proposed in ref.\cite{ugo1}. In ref.\cite{ugo1} the procedure for
computing the slope directly in the HQET was explained. Here we
discuss the computation for QCD, i.e. for heavy quarks with a finite mass.
The advantage of using quarks with a finite mass is that there is
no uncertainty in the normalisation of the lattice operators used to
compute the slope and higher derivatives,
in contrast with the situation in the HQET~\cite{gimenez}.
Moreover the derivatives are computed in terms of ratios of correlation
functions, and it is likely that there will be some cancellation of
systematic errors and statistical fluctuations in these ratios.

In order to determine the Isgur-Wise function it is convenient to
evaluate  the elastic matrix element $\langle P(p^\prime)| \bar Q\gamma^\mu Q
|P(p)\rangle$, where $P(p)$ represents a pseudoscalar meson of mass $m$,
consisting of a heavy quark $Q$ and a light antiquark. This matrix element can
be parametrised by a single form-factor $f(q^2)$ defined by
\beq M^\mu \equiv
\langle P(p^\prime)| \bar Q\gamma^\mu Q |P(p)\rangle
 = (p+p^\prime)^\mu f(q^2) \eeq
where $q=p-p^\prime$ and $q^2=2m^2(1-\omega)$.
The Isgur-Wise function,
$\xi(\omega)$ is given by
\beq f\left( 2m^2(1-\omega)\right) = \left( 1 +
\beta(\omega)\,\right) \xi(\omega ) \ =\ \bar\xi(\omega)
\label{eq:xidef}\eeq where $\beta (\omega)$ represents
radiative corrections which will be discussed later ($\beta(1)=0$ for the
degenerate case we are considering here), and we neglect corrections of
$O(1/m)$.

The derivatives of the Isgur-Wise function at the zero recoil point
can be expressed in terms of the derivatives of $M^\mu$ w.r.t. the momentum
$p$ as will be explained below (in the following we
will take $\mu=4$, which is a convenient choice).
For simplicity consider the case where $p^\prime=(m,\vec 0)$ and
$p=(p_0,0,0,p_3)$, with $p_0=\sqrt{m^2+p_3^2}$
\footnote{In actual calculations we envisage averaging over results
obtained with non-zero components of $\vec p$ in each of the 3 directions.},
so that $M^4$ is a function of $p_3^2$.
The derivatives of $M^4(p_3^2)$ w.r.t. $p_3^2$ are as follows:
\beqn
M^{4\,^\prime}(0) & = & \frac{dM^4}{dp_3^2}|_{p_3^2=0} =
\frac{1}{m}\left(
\bar{\xi}^\prime(1)+\frac{1}{2}\right) \\
M^{4\,^{\prime\prime}}(0) & = & \frac{d^2M^4}{d(p_3^2)^2}|_{p_3^2=0} =
\frac{1}{2m^3}\left(\bar{\xi}^{\prime\prime}(1)
-\frac{1}{2}\right) \eeqn
and similarly
for the higher derivatives \footnote{Throughout this letter we give the
explicit formulae for the first two derivatives.}, where the $^\prime$'s
on $\bar\xi$ denote derivatives w.r.t. $\omega$.

$M^{4\,^\prime}(0)$ and $M^{4\,^{\prime\prime}}(0)$ can be
obtained from suitable combinations of two- and three-point
correlation functions and their derivatives. Let us consider:
\beqn
C_2(t_x;p_3) & = & \sum_{\vec x} e^{ip_3x_3} <0| J_P(x) J_P^{\dagger}(0)|0>
\label{eq:c2def}\\
         & =  &
\frac{Z^2}{2p_0}e^{-p_0t_x}\ +\ \cdots\label{eq:c2asym}\eeqn
and
\beqn
C_3(t_x,t_w; p_3) & = & \sum_{\vec x, \vec w} e^{ip_3 x_3} <0| J_P(w)
V^4(x) J_P^{\dagger}(0)|0>
\label{eq:c3def}\\
         & =  &
\frac{Z^2}{4mp_0}e^{-p_0t_x}e^{-m(t_w-t_x)} M^4(p_3)\ +\ \cdots
\label{eq:c3asym}\eeqn where $J_P$ ($J_P^\dagger$) is an interpolating
operator which can  annihilate (create) the heavy-light pseudoscalar
meson $P$, and $V^4$
represents the fourth component of the vector current
$Z_V\bar Q\gamma^\mu Q$~\footnote{Of course one could also use the lattice
``conserved" vector
current, i.e. the current which is conserved for quarks which are
degenerate in mass.}. $Z_V$ is the renormalisation constant relating the
lattice vector current to the physical one.
$Z=|\langle 0|J_P(0)|P\rangle |$ and
the ellipses in eqs.(\ref{eq:c2asym}) and (\ref{eq:c3asym}) represent terms
which are exponentially suppressed at large time separations, i.e.
when $t_x$ and $t_w-t_x$ are both large. We assume that we can study the
correlation functions at sufficiently large separations to neglect these
terms.
\par In eqs.(\ref{eq:c2def})-(\ref{eq:c3asym}) we have implicitly assumed
that the interpolating operators $J_P$ are local, so that the corresponding
matrix elements $Z$ are independent of $p_3$. In many simulations this
is not the case, extended interpolating operators are used to increase
the overlap with the ground state and the corresponding $Z$'s
depend on momentum. For the
sake of clarity we start our discussion assuming local interpolating operators,
and generalise the results to extended operators towards the end of this
letter.
\par We expand the correlation functions as power series in $p_3$:
\beqn
C_2(t_x;p_3) & = & C_2^{(0)}(t_x) + C_2^{(2)}(t_x)p_3^2 +
C_2^{(4)}(t_x)(p_3^2)^2 + \cdots \\
C_3(t_x,t_w;p_3) & = & C_3^{(0)}(t_x,t_w) + C_3^{(2)}(t_x,t_w)p_3^2 +
C_3^{(4)}(t_x,t_w)(p_3^2)^2 + \cdots
\eeqn
where
\beqn
C_2^{(2)}(t_x) & = & -\frac{1}{2}\sum_{\vec x} x_3^2
\,\langle 0|J_P(x) J^\dagger_P(0)|0\rangle \\
C_2^{(4)}(t_x) & = & \frac{1}{24}\sum_{\vec x} x_3^4
\,\langle 0|J_P(x) J^\dagger_P(0)|0\rangle \\
C_3^{(2)}(t_x,t_w) & = & -\frac{1}{2}\sum_{\vec x,\vec w} x_3^2
\,\langle 0|J_P(w) V^4(x) J^\dagger_P(0)|0\rangle \\
C_3^{(4)}(t_x,t_w) & = & \frac{1}{24}\sum_{\vec x,\vec w} x_3^4
\,\langle 0|J_P(w) V^4(x) J^\dagger_P(0)|0\rangle \eeqn
On a periodic lattice we propose taking $x_3^2=\min\{x_3^2,(L_3-x_3)^2\}$,
where
$L_3$ is the length of the lattice in the $z$-direction, and similarly
$x_3^4=\min\{x_3^4,(L_3-x_3)^4\}$.
The derivatives of the Isgur-Wise function at the zero recoil point are
given simply in terms of the ratios $R_{2,3}^{(2,4)}\equiv C_{2,3}^{(2,4)}/
C_{2,3}^{(0)}$ (where we have suppressed the time labels $t_x$ and $t_w$):
\beqn
\frac{1}{2m^2}\left(\bar{\xi}^\prime(1)
+\frac{1}{2}\right) & = & R_3^{(2)} - R_2^{(2)} \label{eq:xip}\\
\frac{1}{8m^4}\left(\bar{\xi}^{\prime\prime}(1)
-\frac{1}{2}\right) & = &
R_3^{(4)} +(R_2^{(2)})^2 -R_2^{(4)}-R_2^{(2)}R_3^{(2)}\label{eq:xipp}\eeqn
Since $\xi^\prime(1)$ and $\xi^{\prime\prime}(1)$  are obtained from
ratios of correlation functions, it is not necessary to know the
normalisation constant $Z_V$ \footnote{$Z_V$ can however be
determined with excellent precision (to about 0.1\% with $O(100)$
gauge field configurations on typical lattices) using the relation:
$Z_V=C_2(t_x+t_w;0)/C_3(t_x,t_w;0)$.}.
\par In order to obtain the derivatives of the renormalisation group invariant
Isgur-Wise function $\xi(\omega)$ (defined in detail in ref.\cite{neubertprd})
from the results for $\bar\xi^{\,\prime}$
and $\bar\xi^{\,\prime\prime}$,
it is necessary to evaluate the
derivatives of $\beta(\omega)$, the radiative corrections in
eq.(\ref{eq:xidef}). This is a calculation in continuum perturbation theory,
although the results depend on the mass of the heavy quark being
used in the simulations. The ingredients necessary to perform the evaluation
of $\beta^\prime(1)$ and $\beta^{\prime\prime} (1)$ for elastic
matrix elements are presented in the appendix of ref.\cite{neubertprd}.
In the recent simulation by the UKQCD collaboration $m_Q$ was about
1.36 GeV \cite{ukqcd}, for which, taking $\Lambda_{\overline{MS}}=0.25$ GeV,
we find $\beta^\prime (1)\simeq -0.24$ and $\beta^{\prime\prime} (1)\simeq
0.17$
(for $n_f=3$, where $n_f$ is the number of quark flavours in the
evolution equations, however the dependence on the $n_f$ is
mild, e.g. the results are -0.23 and 0.17 for $n_f=0$).
Having obtained the derivatives of $\xi(\omega)$ at the zero recoil point,
it is possible to obtain those of the form-factors
for $B\rightarrow D$ and $D^*$ semi-leptonic decays by using continuum
perturbation theory and the heavy quark symmetry\cite{neubertprd}.
\par The discussion in this letter can readily be generalised to computations
which use extended interpolating operators. In this case the matrix elements
$Z$ become functions of $p_3^2$. Expanding $Z(p_3^2)$ as a power series
in $p_3^2$,
\beq
Z(p_3^2)=Z^{(0)} + Z^{(2)}p_3^2 + Z^{(4)}p_3^4 +\cdots
\eeq
we define $R_Z^{(2,4)}\equiv Z^{(2,4)}/Z^{(0)}$. Equations (\ref{eq:xip})
and (\ref{eq:xipp}) become modified to
\beqn
\frac{1}{2m^2}\left(
\bar{\xi}^\prime(1) + \frac{1}{2}\right) & = & R_3^{(2)} - R_2^{(2)}
+ R_Z^{(2)}\label{eq:xip2}\\
\frac{1}{8m^4}\left(\bar{\xi}^{\prime\prime}(1)
-\frac{1}{2}\right) & = &
R_3^{(4)} +(R_2^{(2)})^2 -R_2^{(4)}-R_2^{(2)}R_3^{(2)}
+ R_Z^{(4)} + \nonumber\\
& & \mbox{\hspace{0.6in}}R_3^{(2)}R_Z^{(2)} - R_2^{(2)}R_Z^{(2)}
\label{eq:xipp2}\eeqn
Thus in order to obtain the derivatives of the Isgur-Wise function using
extended interpolating operators we need to determine $R_Z^{(2)}$ and
$R_Z^{(4)}$. This
is done by fitting $R_2^{(2)}$ and $R_2^{(4)}$ as functions of $t_x$:
\beq
R_2^{(2)} = 2R_Z^{(2)} - \frac{1}{2m^2} -\frac{1}{2m}t_x
\label{eq:rz2}\eeq
and
\beq
R_2^{(4)}
= \frac{1}{8m^2}t_x^2 - \left( \frac{1}{m}R_Z^{(2)}
-\frac{3}{8m^3}\right)t_x + 2R_Z^{(4)}
+\left( R_Z^{(2)}\right) ^2 - \frac{1}{m^2}R_Z^{(2)}
+ \frac{3}{8m^4}\label{eq:rz4}\eeq
\mbox{}\vspace{15pt}
\par In this letter we have proposed a method for the direct determination of
the slope and higher derivatives of the Isgur-Wise function at the zero
recoil point. We expect that this technique, particularly when combined
with standard lattice determinations of $\xi(\omega)$, for discrete values
of $\omega>1$, will provide a valuable tool for the extrapolation of
experimental results to $\omega=1$, and hence will lead to a more
precise determination of $V_{cb}$. The method can also be generalised to
other decays, such as the semileptonic decays of heavy baryons.
\subsection*{Acknowledgements}
We thank L.P.Lellouch, N.Stella and H.Wittig for helpful discussions.
GM acknowledged the partial support of the M.U.R.S.T., Italy and INFN. CTS
acknowledges the support of the Science and Engineering Research Council, UK
through the award of a Senior Fellowship.

\end{document}